\documentclass[preprint]{elsarticle} 
\usepackage{color}
\usepackage{slashed,phonetic}
\usepackage{appendix}

\usepackage{amsmath,latexsym,slashed,dsfont}
\usepackage{mathrsfs}
\usepackage{graphicx} 
\usepackage{amsfonts}
\usepackage{float}
\usepackage{tabularx} 

\newcommand{\beq}{\begin{equation}}
\newcommand{\eeq}{\end{equation}}
\def\bea#1\eea{\begin{align}#1\end{align}}
\def\beal#1\eeal{\begin{subequations}\begin{align}#1\end{align}\end{subequations}}

\newcommand\nn{\nonumber}

\def\f {{\rm \texttt{f}}}

\def\Re           {{\rm Re\hskip0.1em}}
\def\Im           {{\rm Im\hskip0.1em}}

\journal{PLB}

\begin{document}
\begin{frontmatter}
\title{Gauge-Higgs models from Nilmanifolds}

\author{Aldo Deandrea}
\ead{deandrea@ipnl.in2p3.fr}
\author{Fabio Dogliotti} 
\ead{dogliotti@ipnl.in2p3.fr}
\author{Dimitrios Tsimpis}
\ead{tsimpis@ipnl.in2p3.fr}
\address{Universit\'e de Lyon, F-69622 Lyon, France: Universit\'e Lyon 1, Villeurbanne CNRS/IN2P3, UMR5822, Institut de Physique des 2 Infinis de Lyon}

\begin{abstract}
\noindent We consider the compactification of a Yang-Mills theory on a three-dimensional nilmanifold. The compactification generates a Yang-Mills theory  in four space-time dimensions, coupled to a specific scalar sector. The compactification geometry gives rise to masses for the zero-modes, proportional to the twist parameter of the nilmanifold. We study the simple example of an $SU(3)$ model broken by a non-trivial vacuum of the scalar potential which generates three mass scales, two being at tree level, and the third one at loop level. We point out the relevance of general twisted geometries for model building and in particular for gauge-Higgs type models, as the twist generates tree-level mass hierarchies useful for grand unification  and for the Higgs sector in electroweak symmetry breaking.
\end{abstract}

\end{frontmatter}

One of the challenges in model building, when the Higgs potential is radiatively generated, as for example in gauge-Higgs models, is that scalar masses are typically too small, making it difficult to build realistic models for the Higgs sector, both at the unification and the electroweak scales \cite{Manton:1979kb,Hosotani:1983xw,Haba:2004qf,Hosotani:2005nz,Medina:2007hz,Hosotani:2008tx,Hosotani:2015hoa}.
We discuss a new mechanism for generating massive scalar sectors from a partially compactified Yang-Mills theory by deforming the traditional torus-type compact dimension scenario by twisting, which results in the presence of massive scalar fields and a scalar potential at tree level which are completely determined by the gauge structure and by geometry. As an example we use a nilmanifold,  a compact differentiable manifold whose tangent vectors form a nilpotent Lie algebra, see {\it e.g.} \cite{2009arXiv0903.2926B}. Here, we will work with the simplest,  three-dimensional nilmanifold, whose tangent vectors $V_{1,2,3}$ obey  the Heisenberg algebra, 
\bea
\label{heisenberg algebra}
[V_1,V_2]=-\f V_3, \quad [V_1,V_3]=[V_2,V_3]=0~,
\eea
where $\f\in \mathbb{R}$ is the only non-vanishing  structure constant. 
This manifold can be thought of as a twisted fibration of  a circle over a two-dimensional torus.  The twist of the fibration is given by, 
\bea
N = \frac{r_1 r_2}{r_3} \f \text{ \ ,  } N \in \mathbb{Z} ~,
\eea
where $r_{1,2}$ are the two radii of  the torus base, and $r_3$ is the radius of the fiber. 

\par  
Our starting point is a pure Yang-Mills theory with fields valued in  the algebra of a Lie group $G$, 
\beq
\label{action}
{\cal S} = \int_{M_7}\!\! \text{d}^7\! x ~\!{\rm Tr} (t_a t_b)\,  \tfrac12 {\cal F}^a_{MN} {\cal F}^{bMN} \ ,
\eeq
where $\cal F$ is the field strength, 
\bea
 {\cal F}^a_{MN}= 2 \partial_{[M} {\cal A}^a_{N]} +g_{7} f^a{}_{bc} {\cal A}^b_M {\cal A}^c_N  \ ,
\eea
and the $t_a$'s are a basis of the Lie algebra with structure constants $f^a{}_{bc}$.  The total space is $M_7 = {M_4 \times N_3}$, the direct product of Minkowski space and a three-dimensional nilmanifold ${N_3}$;   $M,N$ label  the coordinates of $M_7$. 
The effective four-dimensional action, after truncation of the heavy modes, is \cite{Andriot:2018tmb,Andriot:2020ola}, 
\beq
\int\text{d}^4 \! x ~\!\text{Tr}\Big(
-\tfrac{1}{2} F_{\mu\nu}F^{\mu\nu}+\sum_{I=1}^3D_\mu\Phi_{I} D^\mu\Phi_{I} 
-M^2(\Phi_{3})^2-\mathcal{U}
\Big) ~,
\eeq
where the potential $\mathcal{U}$ reads,
\bea
\mathcal{U}=\text{Tr}\Big(-2i gM[\Phi_1,\Phi_2]\Phi_3 + \tfrac12 g^2\sum_{I,J=1}^3[\Phi_I,\Phi_J][\Phi_I,\Phi_J]\Big)
~.
\eea
The parameters $g$, $M$ are related to the geometry of $N_3$,  
\bea
g = \frac{g_{7}}{\sqrt{V}}~, \quad M = |\f|~,
\eea
where $V=r_1r_2r_3$ is the volume of the nilmanifold. 
The vacuum condition takes the form, 
\bea 
\label{gclass}
\Phi_{3, 0}=0~;~~~
[\Phi_{1, 0},\Phi_{2, 0} ]=0~.
\eea
This class of vacua can accommodate different types of gauge-symmetry breaking patterns. In the following, we take the gauge group to be $SU(3)$, spontaneously broken to $SU(2)\times U(1)$.  Explicitly, the parametrization of the vacuum is given by,
\bea
&\Phi_{1, 0} = \frac{\tilde{b}_1}{\sqrt{3}} \frac{\lambda_8}{2} , \quad \Phi_{2, 0} = \frac{\tilde{b}_2}{\sqrt{3}} \frac{\lambda_8}{2} \text{\quad and \quad}  
\Phi_{3, 0} = 0 ~,
\eea
with $\lambda_8$ the usual $SU(3)$ matrix in Gell-Mann notation.

We expand the scalar fields around their vacuum expectation values (VEV): $\Phi_I = \Phi_{I, 0} + \delta \Phi_I$. 
In order to obtain the mass eigenstates, we define the diagonalization matrix $P$ such that, 
\bea 
\delta \Phi = P(\theta, \nu) \delta \tilde{\Phi}~,
\eea
where $\delta \tilde{\Phi}$ are the scalar fields with definite mass around the vacuum,  written as a 24-component  vector. The parameter  
$\theta \in [ 0,\pi/4 )$ corresponds to the mixing between $\Phi_1$, $\Phi_2$ and $\Phi_3$, while $\nu$ corresponds to the ratio of $\Phi_{1, 0}$ to 
$\Phi_{2, 0}$. The case of $\Phi_{1,0} = \Phi_{2,0}$ is recovered for $\nu = \pi/4$. Explicitly, the relation between $\theta$ and the vacuum parameters is 
\bea
\cos(\theta) & = \sqrt{\frac{1+ \frac{b_1^2+ b_2^2}{4}+\sqrt{1+ \frac{b_1^2+ b_2^2}{2}}}{1+\frac{b_1^2+ b_2^2}{4}+\sqrt{1+\frac{b_1^2+ b_2^2}{2}}}}~,
\eea
while the relation for $\nu$ is 
\bea
\cos(\nu) & = \frac{1}{\sqrt{ \left| \frac{b_1}{b_2} \right| {}^2+1}} ~.
\eea

The Lagrangian in terms of the eigenmass states reads, 
\bea
\mathcal{L}_{dyn} &= -\frac{1}{2}Tr(F_{\mu \nu}F^{\mu \nu})  - D_{\mu}X_{\nu}^{\dagger} \cdot D^{\mu}X^{\nu} + M^{2}_{X} X^{2}\nonumber \\
& + D_{\mu} H_{i}^{\dagger}\cdot D^{\mu} H_{i} - (M_{i})^2 H_{i}^{\dagger}.H_{i} + D_{\mu} H_{0}^{\dagger}\cdot D^{\mu} H_{0} \nonumber\\
& + Tr(D_{\mu} \phi_{i} D^{\mu} \phi_{i}) + Tr(D_{\mu} \rho D^{\mu} \rho) - M_{\rho}^2 Tr(\rho^2)~,
\eea
where $i =1,2$; $F_{\mu \nu}$ is the field strength tensor for the $SU(2)\times U(1)$ gauge; $X_{\mu}$ is the broken boson (gauge-fixed so the dynamical term is no longer antisymmetric in $\mu$, $\nu$); the $H_i$'s, the $\phi_i$'s and $\rho$ are scalars. The masses take the following form, 
\bea
M_{1} &= M f_{-}(x) \quad \quad
M_{2} = M f_{+}(x) \\
M_{\rho} &= M \quad \quad
M_X = M x ~,
\eea
where, 
\bea
x= \frac{\sqrt{b_1^2+b_2^2}}{2}~,
\eea
and we define the function $f_{\pm} (x)$ as, 
\bea
f_{\pm}(x) = \sqrt{(1 + x^2 \pm \sqrt{1+2x^2}) }~.
\eea
In this convention the $b$'s are dimensionless,  so that $b_i = \frac{g}{M}\tilde{b}_i$, and $\tilde{b}_i$ has mass dimension one. All fields of this model transform in either of two $SU(2) \times U(1)$ representations:   the adjoint  ($\phi_i$, $\rho$) and the fundamental ($H_i$, $H_0$, $X_{\mu}$). The adjoint representation   is to be understood as the span of $(\frac{\sqrt{3}}{2}\mathbb{I} ,\frac{\sigma_1}{2}, \frac{\sigma_2}{2}, \frac{\sigma_3}{2})$. The coupling constants are, 
\bea
g_{SU(2)} = \frac{g}{2} \text{\ \ \  and \ \ \ } g_{U(1)} = \frac{\sqrt{3}}{2}g~.
\eea
The boson $X_{\mu}$ becomes massive by acquiring the scalar $H_0$ as longitudinal component. The cubic terms of the scalar potential around the selected vacuum are given by, 
\bea
- 2 g M i Tr ([\phi_1, & \phi_2]\phi_3) = - 2gM\Big(-2\cos(2\theta)\cos(\nu) \Im( H_1^\dagger \cdot \phi_2 \cdot H_2) \nonumber \\ 
&-2\cos(2\theta)\sin(\nu) \Im( H_1^\dagger \cdot \phi_1 \cdot H_2) \nonumber \\
&+ \sin(2\theta)\cos(\nu) H_1^\dagger \cdot \phi_2 \cdot H_1\nn \\ 
&+ \sin(2\theta)\sin(\nu) H_1^\dagger \cdot \phi_1 \cdot H_1\nn \\ 
&- \sin(2\theta)\cos(\nu) H_2^\dagger \cdot \phi_2 \cdot H_2 \nonumber \\
&- \sin(2\theta)\sin(\nu) H_2^\dagger \cdot \phi_1 \cdot H_2 + 2 i Tr(\phi_2 \cdot [\rho , \phi_1])  \Big)~, 
\eea
as well as  the cubic terms coupling the broken boson $X_{\mu}$ to the scalars,
\bea
& - 2g\sin(\theta) \Re \left( \partial_{\mu}H_1^\dagger \cdot \rho \cdot X^{\mu} - H_1^\dagger \cdot \partial_{\mu}\rho \cdot X^{\mu} \right)  \nn \\
& + 2g\cos(\theta) \Im \left( X^{\mu \dagger} \cdot \rho \cdot \partial_{\mu} H_2 - X^{\mu \dagger} \cdot \partial_{\mu}\rho \cdot H_2  \right) \nn \\
& + 2g\cos(\theta) \cos(\nu)\Im \left( X^{\mu \dagger} \cdot \phi_1 \cdot \partial_{\mu} H_1 - X^{\mu \dagger} \cdot \partial_{\mu} \phi_1 \cdot H_1  \right) \nn \\
& -  2g\sin(\theta) \cos(\nu) \Re \left(  \partial_{\mu} H_2^\dagger \cdot \phi_1 \cdot X_{\mu} - H_2^\dagger \cdot \partial_{\mu}\phi_1 \cdot X_{\mu}  \right) \nn \\
& + 2g\cos(\theta) \sin(\nu)\Im \left( X^{\mu \dagger} \cdot \phi_2 \cdot \partial_{\mu} H_1 - X^{\mu \dagger} \cdot \partial_{\mu} \phi_1 \cdot H_1  \right) \nn \\
& -  2g\sin(\theta) \sin(\nu) \Re \left(  \partial_{\mu} H_2^\dagger \cdot \phi_2 \cdot X_{\mu} - H_2^\dagger \cdot \partial_{\mu}\phi_1 \cdot X_{\mu}  \right) \; .
\eea
In addition there is a large number of quartic terms. All the terms are generated by the starting gauge group of the pure Yang-Mills theory in the 7-dimensional theory, once the compatification geometry is specified and the effective, truncated, 4-dimensional theory is explicitly derived.

In the following we are interested in particular in the terms contributing to the mass at one-loop order. Moreover the loops with heavy particles  give dominant contributions, since the diagrams are of order $g^2m^2/16\pi^2$ for a particle of mass  $m$ circulating in the loop. We thus obtain the following terms,
\bea
&g^2Tr([\phi_1, \rho]^2) + 
g^2 Tr([\phi_2, \rho]^2) + 
g^2(1-\sin^2(\theta)\cos^2(\nu)) H_2^{\dagger}\cdot \phi_1^2 \cdot H_2 \nn \\
&+ g^2(1-\sin(\theta)^2\sin^2(\nu)) H_2^{\dagger}\cdot \phi_2^2 \cdot H_2 \nn \\
&+ g^2(1-\cos^2(\theta)\cos^2(\nu)) H_1^{\dagger}\cdot \phi_1^2 \cdot H_1 \nn \\
&+ g^2(1-\cos^2(\theta) \sin^2(\nu)) H_1^{\dagger}\cdot \phi_2^2 \cdot H_1 \nn \\
&+ g^2\cos^2(\theta) H_1^{\dagger}\cdot \rho^2 \cdot H_1 \nn \\
&+ g^2\sin^2(\theta) H_2^{\dagger}\cdot \rho^2 \cdot H_2 \nn \\
&+g^2 X_{\mu}^{\dagger} \cdot \rho^2 \cdot X^{\mu} 
+g^2 \left( ( H_2^{\dagger} \cdot H_2 ) ( X_{\mu}^{\dagger} \cdot X_{\mu} ) -  ( H_2^{\dagger} \cdot \sigma_3 \cdot H_2 ) ( X_{\mu}^{\dagger} \cdot \sigma_3 \cdot X_{\mu} ) \right)  \nn \\
&+ g^2 \left( ( H_1^{\dagger} \cdot H_1 ) ( X_{\mu}^{\dagger} \cdot X_{\mu} ) -  ( H_1^{\dagger} \cdot \sigma_3 \cdot H_1 ) ( X_{\mu}^{\dagger} \cdot \sigma_3 \cdot X_{\mu} ) \right) \nn \\
&+ g^2 \left( ( H_1^{\dagger} \cdot H_1 ) ( H_2^{\dagger} \cdot H_2 ) -  ( H_1^{\dagger} \cdot \sigma_3 \cdot H_1 ) ( H_2^{\dagger} \cdot \sigma_3 \cdot H_2 ) \right)~.
\eea
%
%
%
%
The case we discussed for $SU(3)$ can be easily extended to other gauge groups. One important point for the applications to the electroweak interactions is to obtain a Weinberg angle compatible with data. Since the breaking happens within the adjoint representation, the Weinberg angle of the model is determined by the root diagram of the algebra. Here for example, the embedding of $SU(2) \times U(1)$ in $SU(3)$ is unique and gives a Weinberg angle $\pi/3$. Using the classification of root diagrams, in the case of a rank-2 algebra, a sensible choice could be $G_2$, since it can accommodate a Weinberg angle $\pi/6$, close to the measured value. Of course this group theory value is the one valid at high scale:  to obtain the measured value at the electroweak scale the renormalization group evolution should be taken into account. This is beyond the scope of the present study as a realistic evolution would also require the presence of fermionic matter. Higher-rank algebras allow for more complex breaking patterns but will tend to leave more massless degrees of freedom. These would have to be lifted in mass, or eliminated by other means. 


Limiting ourselves to the case $b_1 = b_2 = b$ for simplicity, and taking $b$ to be small, we can compute the renormalization of the masses at one loop. Of particular interest are the renormalized masses of the light fields: $H_1$, $\phi_1$ and $\phi_2$. The heavy fields will remain at the ``geometrical'' mass scale $M$. Let $\delta M$ denote the one-loop mass corrections. 
We recall that when $b \to 0$ then $\theta \to 0$. The adjoint fields split into  an $SU(2)$ and a $U(1)$ part. This separation becomes important upon renormalization since their couplings, and therefore their renormalized masses, are different.  We denote the fields as,
\bea
\phi_i = \phi_i^{SU(2)} + \phi_i^{U(1)}~,
\eea
with $\phi_i^{SU(2)} \in \text{Span} \{\frac{\sigma_a}{2}\}$ and $\phi_i^{U(1)} \in \text{Span} \{\frac{\sqrt{3}}{2}\mathbb{I} \}$.

\begin{figure}
\centering
\includegraphics[scale= .25]{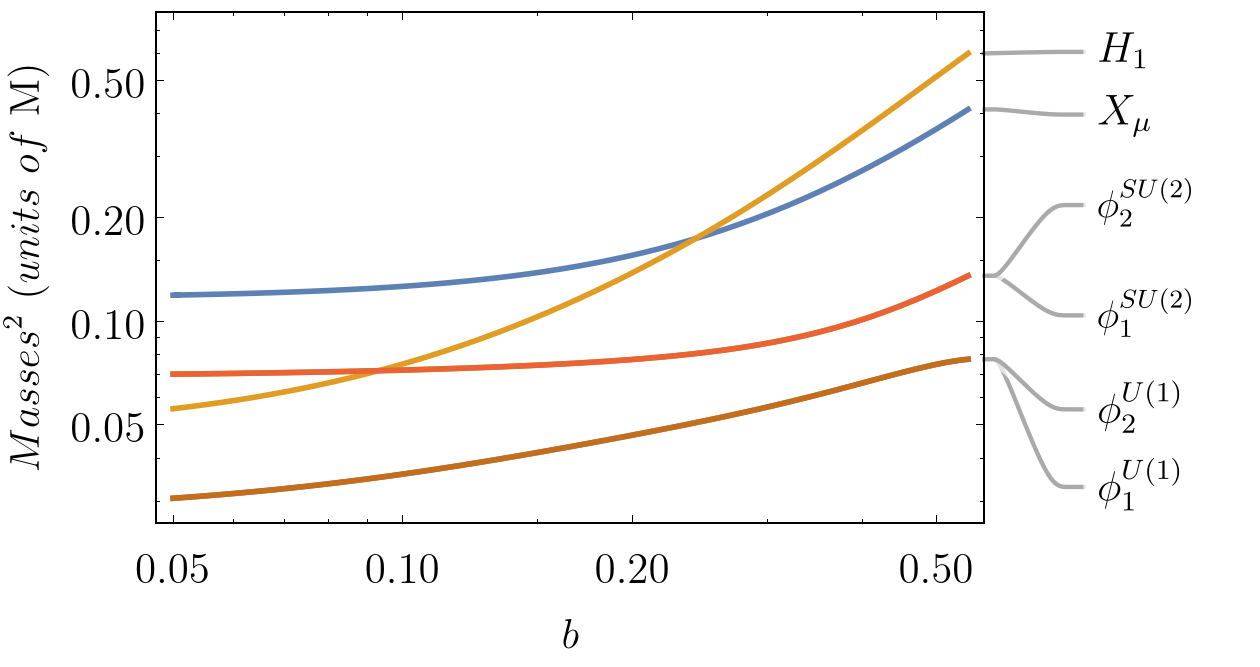}
\caption{Mass spectrum of the theory as a function of the dimensionless vacuum parameter $b$. The masses are in units of $M$, the geometrical mass. The value of $g$ sets the renormalization scale. This plot was computed for a coupling constant $g=10^{-1}$, a UV cutoff $\mu= M$ and the angle $\nu=\pi/4$. The impact of the angle $\nu$ is simply to separate the masses of the states $\phi_1$ and $\phi_2$. The massless scalars are of the same order of magnitude as the Higgs-like field, but couplings to fermions could change the mass hierarchy.}
\label{graph}
\end{figure}

We thus find that the massless scalars aquire a mass of the same order as the Higgs-like field for low vacuum parameter. This mass hierarchy can be modified by the introduction of specific fermion couplings in the model. These results are summarized in Fig. \ref{graph}. 

We have argued that gauge-Higgs unification models based on more general geometries can overcome the shortcomings 
of gauge-Higgs unification models based on simple torus compactifications. We have used the simplest nilmanifold as a concrete example, 
in order to exhibit the generation of mass hierarchies already at tree level. Radiative corrections generate an additional mass scale, while the number of 
scalar species is a function of both the initial grand-unified gauge group and the number of compactified dimensions. 
Even richer mass hierarchies can be generated at tree level, {\it e.g.}~by compactifying on more general group manifolds. Indeed the latter 
are characterized by an underlying Lie algebra with several structure constants, in contrast to the Heisenberg nilmanifold which only has one. 
As we saw, the non-trivial structure constant of the nilmanifold becomes a mass scale from the point of view of the four-dimensional theory. Therefore 
compactifications on more general group manifolds would  lead to several tree-level mass scales. 

The model does not contain any massless  scalars when quantum corrections are taken into account. Our simple model can be thought of as a template for grand-unified models, with the VEV's of the adjoint scalars providing the mechanism for breaking the initial symmetry to the electroweak gauge group. 
The electroweak symmetry breaking mechanism for the Higgs sector is not described in this simple model, but can be included using standard techniques. Without any extra ingredient the model we proposed can spontaneously break the gauge group down to the maximal torus of the gauge group, but the remaining $U(1)$ symmetries cannot be broken without extra ingredients. A well known possibility is the use of orbifolding methods which are known to  provide a further symmetry-breaking mechanism \cite{Hosotani:1988bm,Haba:2002py}. A preliminary investigation of orbifolding of the Heisenberg nilmanifold was initiated in \cite{Andriot:2016rdd}. Besides the gauge and Higgs fields considered here, a complete model must of course also include the fermionic sector,  which could consist of both higher-dimensional fermions and four-dimensional ones (as in the case of brane models). Moreover, the fermionic sector will affect the radiative corrections to the masses, and should be studied in conjunction with the  electroweak symmetry breaking mechanism. These issues will be the subject of future work.

%
%

\bibliography{refs}

\end{document}